\documentclass{gio2}
\usepackage{epsf}

\ifCUPmtlplainloaded \else
  \checkfont{eurm10}
  \iffontfound
    \IfFileExists{upmath.sty}
      {\typeout{^^JFound AMS Euler Roman fonts on the system,
                   using the 'upmath' package.^^J}%
       \usepackage{upmath}}
      {\typeout{^^JFound AMS Euler Roman fonts on the system, but you
                   dont seem to have the}%
       \typeout{'upmath' package installed. JFM.cls can take advantage
                 of these fonts,^^Jif you use 'upmath' package.^^J}%
      }
  \else
  \fi
\fi

\ifCUPmtlplainloaded \else
  \checkfont{msam10}
  \iffontfound
    \IfFileExists{amssymb.sty}
      {\typeout{^^JFound AMS Symbol fonts on the system, using the
                'amssymb' package.^^J}%
       \usepackage{amssymb}%
       \let\le=\leqslant  
         
      }{}
  \fi
\fi

\ifCUPmtlplainloaded \else
  \IfFileExists{amsbsy.sty}
    {\typeout{^^JFound the 'amsbsy' package on the system, using it.^^J}%
     \usepackage{amsbsy}}
    {\providecommand\boldsymbol[1]{\mbox{\boldmath $##1$}}}
\fi

\providecommand\bnabla{\boldsymbol{\nabla}}
\providecommand\bcdot{\boldsymbol{\cdot}}
\providecommand\bcddot{\boldsymbol{:}}
\newcommand\bb{\boldsymbol{b}}
\newcommand\bd{\boldsymbol{d}}
\newcommand\be{\boldsymbol{e}}
\newcommand\bk{\boldsymbol{k}}
\newcommand\bq{\boldsymbol{q}}
\newcommand\br{\boldsymbol{r}}
\newcommand\bu{\boldsymbol{u}}
\newcommand\bx{\boldsymbol{x}}

\newcommand\bB{\boldsymbol{B}}
\newcommand\bC{\boldsymbol{C}}
\newcommand\bF{\boldsymbol{F}}
\newcommand\bJ{\boldsymbol{J}}
\newcommand\bQ{\boldsymbol{Q}}
\newcommand\bU{\boldsymbol{U}}

\newcommand\bOmega{\boldsymbol{\Omega}}

\newcommand\Real{\mbox{Re}} 

\newcommand\zerotensor{\mathsfbi{0}} 
\newcommand\unittensor{\mathsfbi{1}} 
\newcommand\tensorA{\mathsfbi{A}} 
\newcommand\tensorB{\mathsfbi{B}} 
\newcommand\tensorC{\mathsfbi{C}} 
\newcommand\maxwell{\mathsfbi{M}} 
\newcommand\tensorQ{\mathsfbi{Q}} 
\newcommand\shear{\mathsfbi{S}} 
\newcommand\stress{\mathsfbi{T}} 
\newcommand\tensorX{\mathsfbi{X}} 

\title[Viscoelastic and magnetohydrodynamic flows and their
instabilities] {On the relation between viscoelastic and
  magnetohydrodynamic flows and their instabilities}

\author[G. I. Ogilvie and M. R. E. Proctor] {G\ls O\ls R\ls D\ls O\ls
  N\ns I.\ns O\ls G\ls I\ls L\ls V\ls I\ls E$^{1,2}$ \and M\ls I\ls
  C\ls H\ls A\ls E\ls L\ns R.\ns E.\ns P\ls R\ls O\ls C\ls T\ls O\ls
  R$^2$}

\affiliation{$^1$Institute of Astronomy, University of Cambridge,
  Madingley Road, Cambridge CB3 0HA, UK\\[\affilskip]
  $^2$Department of Applied Mathematics and Theoretical Physics,
  University of Cambridge, Centre for Mathematical Sciences,
  Wilberforce Road, Cambridge CB3 0WA, UK}

\begin{document}

\maketitle

\begin{abstract}
  We demonstrate a close analogy between a viscoelastic medium and an
  electrically conducting fluid containing a magnetic field.
  Specifically, the dynamics of the Oldroyd-B fluid in the limit of
  large Deborah number corresponds to that of a magnetohydrodynamic
  (MHD) fluid in the limit of large magnetic Reynolds number.  As a
  definite example of this analogy, we compare the stability
  properties of differentially rotating viscoelastic and MHD flows.
  We show that there is an instability of the Oldroyd-B fluid that is
  physically distinct from both the inertial and elastic instabilities
  described previously in the literature, but is directly equivalent
  to the magnetorotational instability in MHD.  It occurs even when
  the specific angular momentum increases outwards, provided that the
  angular velocity decreases outwards; it derives from the kinetic
  energy of the shear flow and does not depend on the curvature of the
  streamlines.  However, we argue that the elastic instability of
  viscoelastic Couette flow has no direct equivalent in MHD.
\end{abstract}

\section{Introduction}

\subsection{Viscoelastic and magnetohydrodynamic fluids}

In his investigation of the viscosity of gases, Clerk Maxwell (1867)
proposed that the stress in a fluid obeys an equation of the form
\begin{equation}
  (\mbox{stress})+\tau{{{\rm d}(\mbox{stress})}\over{{\rm d}t}}=
  (\mbox{viscosity})\times{{{\rm d}(\mbox{strain})}\over{{\rm d}t}},
  \label{maxwell}
\end{equation}
where $\tau$ is the relaxation time.  If the time scale of the
straining motion is long compared to $\tau$, the second term on the
left-hand side is negligible and the stress is proportional to the
rate of strain.  This Newtonian relation gives rise to viscous
behaviour.  However, if the time scale of the strain is short compared
to $\tau$, the first term on the left-hand side is negligible and the
stress is proportional to the strain itself.  This Hookean relation
gives rise to elastic behaviour.  The reason for the elastic response
is that the rapid strain prevents the configuration of the molecules
from relaxing towards an equilibrium distribution, and the stress is
therefore `frozen in' to the fluid.

Modern constitutive equations for viscoelastic fluids (Bird, Armstrong
\& Hassager 1987{\it a\/}) are usually expressed in a covariant
tensorial form based on the principles set out by Oldroyd (1950).  His
liquid B is one of the most widely used nonlinear models of a
viscoelastic fluid, and provides a fair representation of a dilute
solution of a polymer of high molecular weight.  It is based on
Maxwell's equation (\ref{maxwell}), but cast in a form that satisfies
the principle of material frame indifference.  Moreover, it can also
be derived from the kinetic theory of idealized extensible polymer
molecules contained in a Newtonian solvent (Bird, Curtiss \& Armstrong
1987{\it b\/}).  The dimensionless number characterizing the ratio of
the relaxation time to the time scale of the flow is the Deborah
number (or Weissenberg number); when this is large, the polymeric
stress is effectively `frozen in' to the fluid.

In an electrically conducting fluid, the magnetic field $\bB$ affects
the dynamics through the bulk Lorentz force (e.g. Roberts 1967).  This
can be represented in terms of the Maxwell electromagnetic stress
tensor,\footnote{Here $\mu_0$ is the permeability of free space.  For
  non-relativistic flows, the electric field makes a negligible
  contribution to the stress.}
\begin{equation}
  \maxwell={{\bB\bB}\over{\mu_0}}-{{B^2}\over{2\mu_0}}\unittensor,
  \label{maxwell_stress}
\end{equation}
the two parts of which correspond to a tension in the field lines and
an isotropic magnetic pressure.  It is well known that, in a perfectly
conducting fluid, the magnetic field is `frozen in' to the fluid, in
the sense that magnetic field lines can be identified with material
lines (Alfv\'en 1950).  Even in a fluid of finite conductivity, the
magnetic field is effectively `frozen in' for motions of sufficiently
short time scale, or sufficiently large length scale, corresponding to
a large magnetic Reynolds number.  It follows that the Maxwell stress
is also `frozen in' to the fluid in a certain sense.

From a mathematical point of view, the `freezing in' of a tensor field
$\tensorX(\br,t)$ in a flow with velocity field $\bu(\br,t)$ can be
expressed by the equation (e.g. Tur \& Yanovsky 1993)
\begin{equation}
  {{\partial\tensorX}\over{\partial t}}+\pounds_{\bu}\tensorX=\zerotensor,
\end{equation}
where $\pounds$ is the Lie derivative.  For a scalar field $X$, this
gives the familiar expression $\partial X/\partial t+\bu\bcdot\bnabla
X=0$, meaning that the numerical value of $X$ is conserved by every
fluid element.  For a (contravariant) vector field $\bB$ it gives the
induction equation of ideal, incompressible magnetohydrodynamics
(MHD), which implies that the magnetic field is advected and stretched
in the same way as infinitesimal line elements.  For a second-rank
tensor field it results in the `upper-convected derivative' that
appears in the governing equation of the Oldroyd-B fluid (Bird {\it et
  al.}  1987{\it a\/}).

At a physical level, an analogy is to be seen between a polymer
solution, containing extensible molecules that are advected and
distorted by the flow and react on it through their tension, and an
electrically conducting fluid, containing magnetic field lines that
are also advected and distorted by the flow and react on it through
their tension.

It follows that there is a physical and mathematical similarity
between the dynamics of viscoelastic and MHD fluids.  We will show
that a formal analogy can be drawn between the Oldroyd-B fluid in the
limit of large Deborah number and an MHD fluid in the limit of large
magnetic Reynolds number.  In other words, in this limit, the Maxwell
stress in MHD obeys the equation of a Maxwell fluid.

\subsection{Instabilities of differentially rotating fluids}

\label{instabilities}

Differentially rotating flows are common in astrophysics and
geophysics, and have been studied extensively in the laboratory.  The
simplest form of differential rotation occurs when the angular
velocity depends only on the cylindrical radius, $\Omega=\Omega(r)$,
and Couette flow between differentially rotating cylinders provides an
excellent model system for investigating the dynamics of such flows.
According to Rayleigh (1916), instability occurs in the absence of
viscosity whenever the specific angular momentum $|r^2\Omega|$
decreases outwards.  Most subsequent theoretical and experimental
studies, starting with the classic work of Taylor (1923), have focused
on the onset of Rayleigh's inertial instability in the presence of
viscosity, and the interesting sequence of dynamical states that
ensues.

Numerous variants of Couette flow have also been considered, and among
these the Couette flow of viscoelastic fluids has received much
attention.  Early work in the 1960s (e.g. Thomas \& Walters 1964;
Giesekus 1966) examined the effect of viscoelasticity on the onset of
Rayleigh's inertial instability.  However, one of the most important
recent results is the theoretical and experimental demonstration by
Larson, Shaqfeh \& Muller (1990) of a physically distinct instability
in viscoelastic Couette flow.  This is a purely elastic instability
that occurs at sufficiently large Deborah number $\tau|{\rm
  d}\Omega/{\rm d}\ln r|$, even in the limit of negligible inertia,
and irrespective of the sign of the angular momentum gradient.

The influence of a magnetic field on the stability of the Couette flow
of an electrically conducting fluid has also been investigated
theoretically and (to a lesser extent) experimentally.  Again, early
work (described by Chandrasekhar 1961) focused on the effect of the
magnetic field on the onset of inertial instability.  More
importantly, Velikhov (1959) and Chandrasekhar (1960) uncovered a
physically distinct instability in magnetized Couette flow.  In the
absence of viscosity and resistivity, and in the presence of a weak
vertical magnetic field, this `magnetorotational' instability occurs
whenever the angular velocity $|\Omega|$ decreases outwards,
irrespective of the sign of the angular momentum gradient.

The magnetorotational instability finds its most important
applications in astrophysical fluid dynamics, where magnetic fields
are prevalent and the astronomical length scales allow for large
magnetic Reynolds numbers.  The stability of differentially rotating
flows is of considerable interest in astrophysics, especially in
connection with accretion discs (e.g. Pringle 1981).  These are
usually thin discs of gas in circular orbital motion around a star or
black hole.  The angular velocity decreases outwards according to
Kepler's third law, $\Omega\propto r^{-3/2}$, and the Reynolds numbers
are extremely high (e.g. $10^{14}$).  Observations indicate that
angular momentum is transported outwards through accretion discs at a
much greater rate than allowed by viscosity, and understanding the
origin of this `anomalous viscosity' has been a major goal of
accretion disc research.

The anomalous viscosity is usually attributed to turbulent transport.
However, despite the very high Reynolds numbers, there is no
convincing demonstration of any suitable hydrodynamic instability in
circular Keplerian flow.  Indeed, simple reasoning can be used to
argue that hydrodynamic turbulence is unlikely to be self-sustaining
in a flow that amply satisfies Rayleigh's stability criterion (Balbus
\& Hawley 1998).  However, it has been demonstrated that MHD
turbulence develops very readily in accretion discs, through the
nonlinear development of the magnetorotational instability.  Since the
results of Velikhov (1959) and Chandrasekhar (1960) were rediscovered
by Balbus \& Hawley (1991) and their significance was appreciated, the
magnetorotational instability has been analysed in considerable detail
in the astrophysical literature.

\subsection{Properties of the magnetorotational instability}

\label{mri}

The properties of the magnetorotational instability have been reviewed
by Balbus \& Hawley (1998), and we recall some of the important
features here.  Its simplest manifestation is in an incompressible,
inviscid, perfectly conducting fluid having angular velocity
$\Omega(r)$ and containing a uniform magnetic field $\bB$ parallel to
the axis of rotation.  An approximate local dispersion relation can be
obtained for normal modes having growth rate $s$ and wavevector $\bk$
parallel to $\bB$, and has the form
\begin{equation}
  s^4+s^2\left[4\Omega(\Omega-A)+2\omega_{\rm A}^2\right]+
  \omega_{\rm A}^2(\omega_{\rm A}^2-4\Omega A)=0,
  \label{magnetorotational_dispersion_relation}
\end{equation}
where the quantity
\begin{equation}
  A=-{{r}\over{2}}{{{\rm d}\Omega}\over{{\rm d}r}}
\end{equation}
measures the differential rotation, and is known as Oort's first
constant in the astrophysical literature.  The Alfv\'en frequency is
$\omega_{\rm A}=(\mu_0\rho)^{-1/2}\bk\bcdot\bB$, and the combination
\begin{equation}
  4\Omega(\Omega-A)={{1}\over{r^3}}{{\rm d}\over{{\rm d}r}}(r^4\Omega^2)
\end{equation}
is the Rayleigh discriminant, or the square of the epicyclic
oscillation frequency.  When this is positive, unstable normal modes
with $s>0$ can nevertheless be found provided that $4\Omega A>0$, as
is the case in astrophysical discs.  The maximal growth rate, $s=|A|$,
is achieved by a mode having $\omega_{\rm A}^2=A(2\Omega-A)$.

More generally, and from a local perspective, rotation and shear in
the correct relative orientation are required, and a weak magnetic
field of any geometry is sufficient to initiate the instability,
provided the fluid is sufficiently ionized.  In the presence of
significant dissipation, the ideal growth rate must compete with
viscous and resistive damping, so that growth rates less than $|A|$
are achieved, or the instability may be suppressed altogether.  An
unstable mode must always bend the field lines, having a non-zero
Alfv\'en frequency, and therefore the instability of a purely
azimuthal (or toroidal) field is essentially non-axisymmetric.

A simple explanation of the instability can be given in terms of two
fluid elements, connected by magnetic field lines, that are initially
in circular orbit at the same radius.  The fluid elements are then
given angular momentum perturbations of opposite sign.  The one
receiving the positive perturbation moves to an orbit of larger radius
and acquires a smaller angular velocity, lagging behind its partner.
The tension of the magnetic field exerts a torque that pulls the
lagging element forwards, enhancing the initial perturbation and
leading to instability.  A mechanical analogue, consisting of two
orbiting particles connected by a weak spring, also exhibits
instability.

\subsection{Plan of the paper}

The main purpose of this paper is to draw attention to the physical
and mathematical similarity between viscoelasticity and MHD.  As an
example of the application of this idea, we explore in some detail the
relation between the instabilities of differentially rotating
viscoelastic and MHD flows.  As described in \S~\ref{instabilities},
previous investigations have uncovered instabilities of viscoelastic
and MHD Couette flow that are physically distinct from Rayleigh's
inertial instability.  In the light of the analogy we describe, an
obvious question is whether the elastic instability of Larson {\it et
  al.} (1990) is somehow related to the magnetorotational instability.
We will argue that this is not the case, but will show that there is
another instability of viscoelastic Couette flow that is the direct
equivalent of the magnetorotational instability.

The remainder of this paper is organized as follows.  In \S~2 we set
out the basic equations governing incompressible viscoelastic and MHD
flows and present the analogy between them.  We then discuss, in \S~3,
the possible sources of instability on the basic of energy
considerations.  In \S~4 we define a model system consisting of plane
Couette flow in a rotating channel, equivalent to cylindrical Couette
flow in the narrow-gap limit, and formulate eigenvalue problems for
the normal modes of the Oldroyd-B and MHD fluids.  Some numerical
solutions are presented in \S~5 to illustrate the expected similarity
between the two systems.  In \S~6 the existence of localized growing
solutions in the two systems, satisfying the same magnetorotational
dispersion relation, is demonstrated by an asymptotic analysis.
Finally, the results are summarized and discussed in \S~7.

\section{Basic equations}

\subsection{Viscoelastic fluid}

\label{basic_viscoelastic}

We first consider an incompressible viscoelastic fluid of uniform
density $\rho$.  The Oldroyd-B model is characterized by a solvent
viscosity $\mu$, a polymer viscosity $\mu_{\rm p}$ and a relaxation
time $\tau$.  The velocity field $\bU$ obeys the solenoidal condition,
\begin{equation}
  \bnabla\bcdot\bU=0,
\end{equation}
and the equation of motion
\begin{equation}
  \rho\left({{\partial\bU}\over{\partial t}}+\bU\bcdot\bnabla\bU\right)=
  -\bnabla\Psi+\bnabla\bcdot\stress+\mu\nabla^2\bU.
\end{equation}
Here $\Psi=p+\rho\Phi$ is the modified pressure ($\Phi$ being the
gravitational potential) and $\stress$ is the Oldroyd-B stress, which
is a symmetric tensor field of second rank satisfying the constitutive
equation (cf. equation (\ref{maxwell}))
\begin{equation}
  \stress+\tau\left[{{\partial\stress}\over{\partial t}}+
  \bU\bcdot\bnabla\stress-(\bnabla\bU)^{\rm T}\bcdot\stress-
  \stress\bcdot\bnabla\bU\right]=
  \mu_{\rm p}\left[\bnabla\bU+(\bnabla\bU)^{\rm T}\right],
  \label{stress}
\end{equation}
where the superscript `T' denotes the transpose of a second-rank
tensor.

\subsection{MHD fluid}

\label{basic_mhd}

We also consider an incompressible, electrically conducting fluid of
uniform density $\rho$, viscosity $\mu$ and electrical conductivity
$\sigma$.  The velocity field $\bU$ obeys the solenoidal condition,
\begin{equation}
  \bnabla\bcdot\bU=0,
\end{equation}
and the equation of motion
\begin{equation}
  \rho\left({{\partial\bU}\over{\partial t}}+\bU\bcdot\bnabla\bU\right)=
  -\bnabla\Psi+\bnabla\bcdot\maxwell+\mu\nabla^2\bU.
\end{equation}
Here $\Psi=p+\rho\Phi$ is again the modified pressure and $\maxwell$
is the Maxwell stress given in equation (\ref{maxwell_stress}).  The
magnetic field obeys the solenoidal condition,
\begin{equation}
  \bnabla\bcdot\bB=0,
\end{equation}
and the induction equation,
\begin{equation}
  {{\partial\bB}\over{\partial t}}+\bU\bcdot\bnabla\bB=
  \bB\bcdot\bnabla\bU+\eta\nabla^2\bB,
  \label{induction}
\end{equation}
which is derived from Maxwell's equations for the electromagnetic
field and Ohm's law.  Here $\eta=1/(\mu_0\sigma)$ is the magnetic
diffusivity.

\subsection{The formal analogy}

Instead of comparing the Oldroyd-B stress $\stress$ and the Maxwell
stress $\maxwell$ directly, we take the polymeric and magnetic
stress tensors to be
\begin{equation}
  \stress_{\rm p}=\stress+{{\mu_{\rm p}}\over{\tau}}\unittensor,\qquad\qquad
  \stress_{\rm m}={{\bB\bB}\over{\mu_0}},
  \label{tptm}
\end{equation}
which differ from $\stress$ and $\maxwell$ only by the addition of an
isotropic part in each case.  Specifically, $\stress_{\rm p}$ does not
include the equilibrium isotropic pressure $nkT$ of the polymer
molecules, and $\stress_{\rm m}$ does not include the isotropic
pressure $B^2/2\mu_0$ of the magnetic field.  In an incompressible
fluid, such terms can be taken care of by writing the two equations of
motion in the form
\begin{equation}
  \rho\left({{\partial\bU}\over{\partial t}}+\bU\bcdot\bnabla\bU\right)=
  -\bnabla\Psi_{\rm p,m}+\bnabla\bcdot\stress_{\rm p,m}+\mu\nabla^2\bU,
\end{equation}
where
\begin{equation}
  \Psi_{\rm p}=\Psi+{{\mu_{\rm p}}\over{\tau}},\qquad\qquad
  \Psi_{\rm m}=\Psi+{{B^2}\over{2\mu_0}}
\end{equation}
are redefined modified pressures.

According to the Oldroyd-B constitutive equation (\ref{stress}) and
the induction equation (\ref{induction}), $\stress_{\rm p}$ and
$\stress_{\rm m}$ satisfy the equations
\begin{equation}
  {{\partial\stress_{\rm p}}\over{\partial t}}+
  \bU\bcdot\bnabla\stress_{\rm p}-
  (\bnabla\bU)^{\rm T}\bcdot\stress_{\rm p}-
  \stress_{\rm p}\bcdot\bnabla\bU=
  -{{1}\over{\tau}}\left(\stress_{\rm p}-
  {{\mu_{\rm p}}\over{\tau}}\unittensor\right),
  \label{tp}
\end{equation}
\begin{equation}
  {{\partial\stress}_{\rm m}\over{\partial t}}+
  \bU\bcdot\bnabla\stress_{\rm m}-
  (\bnabla\bU)^{\rm T}\bcdot\stress_{\rm m}-
  \stress_{\rm m}\bcdot\bnabla\bU={{\eta}\over{\mu_0}}
  \left[\bB\nabla^2\bB+(\nabla^2\bB)\bB\right].
  \label{tm}
\end{equation}
In the limits $\tau\to\infty$ and $\eta\to0$, corresponding to large
Deborah number and large magnetic Reynolds number respectively, the
right-hand sides of these equations are negligible.  The polymeric and
magnetic stresses then satisfy identical equations, involving the same
upper-convected derivative, and they appear identically in the
equation of motion of the fluid.  Therefore the formal analogy can be
expressed symbolically as
\begin{equation}
  \lim_{\tau\to\infty}(\mbox{Oldroyd-B fluid})=
  \lim_{\eta\to0}(\mbox{MHD fluid}).
\end{equation}

We note that $\stress_{\rm m}$ is a positive semi-definite tensor
having one non-negative eigenvalue and two zero eigenvalues.  Joseph
(1990) has shown that $\stress_{\rm p}$ also retains a positive
definite character when it evolves according to equation (\ref{tp}).
This is required on physical grounds, because in the derivation of the
Oldroyd-B constitutive equation from kinetic theory, $\stress_{\rm
  p}\propto\langle\bd\bd\rangle$, where $\bd$ is the separation of the
ends of a polymer molecule, and the angle brackets denote an average
(Bird {\it et al.} 1987{\it b\/}).  Equation (\ref{tp}) shows that
$\stress_{\rm p}$ attempts to return to isotropy on the relaxation
time, but in a shear flow at large Deborah number, this tendency is
overcome and one eigenvalue of $\stress_{\rm p}$ does indeed dominate,
as required by the MHD analogy.

As we have shown, the induction equation of ideal MHD provides an
equation for the magnetic stress tensor, which is comparable to the
constitutive equation of the Oldroyd-B fluid.  One might ask whether
the constitutive equation can be reduced to something resembling an
induction equation.  This is indeed so: at any instant we may express
the positive definite tensor $\stress_{\rm p}$ in terms of {\it
  three\/} vector fields $\bB_i$,
\begin{equation}
  \stress_{\rm p}={{1}\over{\mu_0}}\sum_{i=1}^3\bB_i\bB_i.
  \label{bb}
\end{equation}
Equation (\ref{tp}) is recovered if the fields evolve according to the
induction-like equations
\begin{equation}
  {{\partial\bB_i}\over{\partial t}}+\bU\bcdot\bnabla\bB_i=
  \bB_i\bcdot\bnabla\bU-{{1}\over{2\tau}}
  \left(\bB_i-{{\mu_0\mu_{\rm p}}\over{\tau}}\bQ_i\right),
  \label{q}
\end{equation}
provided that the fields $\bQ_i$ satisfy
\begin{equation}
  {{1}\over{2}}\sum_{i=1}^3(\bB_i\bQ_i+\bQ_i\bB_i)=\unittensor.
  \label{bq}
\end{equation}
To find the fields $\bQ_i$, let $\tensorB$ be the matrix whose columns
are $(\bB_1,\bB_2,\bB_3)$, and similarly for $\tensorQ$.  In matrix
notation, equation (\ref{bq}) reads
\begin{equation}
  {\textstyle{{1}\over{2}}}(\tensorB\tensorQ^{\rm T}+
  \tensorQ\tensorB^{\rm T})=\unittensor
\end{equation}
and is satisfied when
\begin{equation}
  \tensorQ=(\unittensor+\tensorA)\tensorC,
\end{equation}
where $\tensorA$ is an arbitrary antisymmetric matrix and $\tensorC$
is the inverse of $\tensorB^{\rm T}$.  This means that
\begin{equation}
  \bQ_i=\bC_i+\bOmega\times\bC_i,
\end{equation}
where $\bOmega$ is an arbitrary vector field.  The vector fields
$\bC_i$ are just the reciprocal vectors to $\{\bB_i\}$, e.g.
\begin{equation}
  \bC_1={{\bB_2\times\bB_3}\over{\bB_1\bcdot(\bB_2\times\bB_3)}}.
\end{equation}

The non-uniqueness of the fields $\bQ_i$ reflects the fact that the
representation (\ref{bb}) is partially redundant: we are expressing a
tensor field with six independent components in terms of three vector
fields each having three independent components.  It is therefore
permissible to impose three constraints on the fields $\bB_i$, and
then $\bOmega$ will no longer be arbitrary.  For example, it may be
convenient to require that the fields $\bB_i$ be solenoidal.  Now
equation (\ref{q}) implies that
\begin{equation}
  \left({{\partial}\over{\partial t}}+\bU\bcdot\bnabla\right)
  (\bnabla\bcdot\bB_i)=-{{1}\over{2\tau}}\left(\bnabla\bcdot\bB_i-
  {{\mu_0\mu_{\rm p}}\over{\tau}}\bnabla\bcdot\bQ_i\right),
\end{equation}
and so the solenoidal property is preserved if $\bOmega$ is chosen
such that
\begin{equation}
  0=\bnabla\bcdot\bQ_i=\bnabla\bcdot(\bC_i+\bOmega\times\bC_i),
  \qquad i=1,2,3.
\end{equation}
Choosing the fields $\bB_i$ to be solenoidal also ensures that
\begin{equation}
  \bnabla\bcdot\stress_{\rm p}=
  {{1}\over{\mu_0}}\sum_{i=1}^3\bB_i\bcdot\bnabla\bB_i,
\end{equation}
for direct comparability with the Lorentz force.

\section{Energetics and instability}

\label{energetics}

\subsection{MHD fluid}

Some insight into the possible instabilities of viscoelastic and MHD
flows can be obtained on the basis of energy considerations.
Instabilities typically release energy stored in the basic state and
use this to allow a perturbation to grow in time.  This restricts the
class of flows that can exhibit instability, and limits the growth
rates that can be achieved.

We start by considering the case of the MHD fluid, which is more
straightforward.  Starting from the equations of \S~\ref{basic_mhd} it
is possible to derive an energy equation of the form
\begin{equation}
  {{\partial E}\over{\partial t}}+\bnabla\bcdot\bF=-D,
\end{equation}
where
\begin{equation}
  E={\textstyle{{1}\over{2}}}\rho U^2+{{B^2}\over{2\mu_0}}
\end{equation}
is the energy density,
\begin{equation}
  \bF=(E+\Psi)\bU-{{1}\over{\mu_0}}(\bB\bcdot\bU)\bB-
  \mu\bU\times(\bnabla\times\bU)-
  {{\eta}\over{\mu_0}}\bB\times(\bnabla\times\bB)
\end{equation}
is the energy flux, and
\begin{equation}
  D=\mu|\bnabla\times\bU|^2+{{\eta}\over{\mu_0}}|\bnabla\times\bB|^2
\end{equation}
is the dissipation rate.

Consider a perturbative solution of the equations of
\S~\ref{basic_mhd} in which upper-case symbols $(\bU,\bB,\Psi)$ denote
the basic state (not necessarily steady) and lower-case symbols
$(\bu,\bb,\psi)$ denote the Eulerian perturbations.  Using the
linearized equations, it is then possible to derive the energy-like
equation
\begin{eqnarray}
  {{\partial}\over{\partial t}}\left({\textstyle{{1}\over{2}}}\rho u^2+
  {{b^2}\over{2\mu_0}}\right)+\bnabla\bcdot\bF'&=&
  \left(-\rho\bu\bu+{{\bb\bb}\over{\mu_0}}\right)
  \bcddot\bnabla\bU+(\bb\times\bu)\bcdot\bJ\nonumber\\
  &&\qquad-\mu\left|\bnabla\times\bu\right|^2-
  {{\eta}\over{\mu_0}}\left|\bnabla\times\bb\right|^2
  \label{energy-like}
\end{eqnarray}
governing the perturbations, where $\bF'$ is a certain flux and
$\bJ=\mu_0^{-1}\bnabla\times\bB$ is the current density in the basic
state.  The quantity differentiated with respect to time is the part
of the energy density at second order in the perturbation amplitude
that must grow in any instability.  Provided that the instability is
local, so that it does not depend on a particular choice of boundary
conditions, the term $\bnabla\bcdot\bF'$ cannot play an essential role
in this equation, because it will vanish on integration over the
volume of the fluid in the case of periodic boundary conditions or, in
many cases, physical boundary conditions.  Therefore any local
instability must derive its energy either from the kinetic energy of
the basic flow, through the term involving $\bnabla\bU$, or from the
magnetic energy, through the term involving $\bJ$.  For kinetic energy
to be released, there must be a velocity {\it gradient\/}, because a
uniform flow can be eliminated by a Galilean transformation and
therefore cannot be a source of instability.  A potential magnetic
field ($\bJ={\bf0}$) also cannot be a source of instability, as it
minimizes the magnetic energy in a region subject to the magnetic flux
through its boundary being prescribed (e.g. Priest 1982).

In the case of a potential magnetic field it is possible to place an
upper bound on the growth rate of any local instability.  Let
\begin{equation}
  \shear={\textstyle{{1}\over{2}}}\left[\bnabla\bU+
  (\bnabla\bU)^{\rm T}\right]
\end{equation}
be the rate-of-strain tensor of the basic flow, and
$(\lambda_1,\lambda_2,\lambda_3)$ its eigenvalues.  Its quadratic form
satisfies the inequalities
\begin{equation}
  \min(\lambda_1,\lambda_2,\lambda_3)\le
  {{\shear\bcddot\bx\bx}\over{x^2}}\le\max(\lambda_1,\lambda_2,\lambda_3)
\end{equation}
and therefore
\begin{equation}
  {{|\shear\bcddot\bx\bx|}\over{x^2}}\le
  \max(|\lambda_1|,|\lambda_2|,|\lambda_3|).
\end{equation}
It follows from equation (\ref{energy-like}) that the largest possible
growth rate of any local instability is the largest eigenvalue, in
absolute value, of the rate-of-strain tensor of the basic flow.  When
the magnetic field is not potential the growth rate can be increased
by at most $(\mu_0/\rho)^{1/2}|\bJ|/2$.

\subsection{Viscoelastic fluid}

By working with the representation (\ref{bb}) of the polymeric stress
in terms of three vector fields $\bB_i$ it is possible to derive a
similar energy-like equation
\begin{eqnarray}
  &&{{\partial}\over{\partial t}}\left({\textstyle{{1}\over{2}}}\rho u^2+
  \sum_{i=1}^3{{b_i^2}\over{2\mu_0}}\right)+\bnabla\bcdot\bF''=
  \left(-\rho\bu\bu+\sum_{i=1}^3{{\bb_i\bb_i}\over{\mu_0}}\right)
  \bcddot\bnabla\bU+\sum_{i=1}^3(\bb_i\times\bu)\bcdot\bJ_i\nonumber\\
  &&\qquad+{{1}\over{\mu_0}}\sum_{i=1}^3(\bnabla\bcdot\bb_i)\bu\bcdot\bB_i-
  \mu\left|\bnabla\times\bu\right|^2-
  {{1}\over{\tau}}\sum_{i=1}^3{{b_i^2}\over{2\mu_0}}+
  {{\mu_{\rm p}}\over{2\tau^2}}\sum_{i=1}^3\bb_i\bcdot\bq_i
\end{eqnarray}
governing linear perturbations from any basic state, where $\bF''$ is
a certain flux and $\bJ_i=\mu_0^{-1}\bnabla\times\bB_i$ by analogy
with MHD.  If we constrain the representation such that the fields are
solenoidal, then the term involving $\bnabla\bcdot\bb_i$ vanishes.
The argument proceeds almost as before, with gradients in the basic
flow or the basic stress providing potential sources of energy for the
disturbance.  The final term, involving $\bb_i\bcdot\bq_i$, is a third
possible source of energy, but the $\tau^{-2}$ dependence suggests
that the effect of this term may be expected to be small in the limit
of large Deborah number.

\section{Plane Couette flow in a rotating channel}

As a minimal model of a differentially rotating flow, we consider a
linear shear flow (plane Couette flow) in a rotating channel.  This is
equivalent to cylindrical Couette flow in the limit of a narrow gap,
if the angular velocities of the two cylinders are not widely
disparate.  All effects of curvature are then neglected.

\subsection{Basic state and boundary conditions}

We adopt Cartesian coordinates $(x,y,z)$ in a frame of reference
rotating with uniform angular velocity $\Omega\,\be_z$.  The only
change required to the equations in the rotating frame is the
inclusion of the Coriolis force.  The centrifugal force, which is
derivable from a potential, can be absorbed into the modified
pressure.  The rotation of the frame does not affect any of the other
equations.  The equation of motion therefore becomes
\begin{equation}
  \rho\left({{\partial\bU}\over{\partial t}}+\bU\bcdot\bnabla\bU
  +2\Omega\,\be_z\times\bU\right)=\cdots.
\end{equation}

We consider flow in the channel $0<x<d$ between a stationary plane
boundary $x=0$ and a moving plane boundary $x=d$ with velocity
$-2Ad\,\be_y$.  The non-slip and impermeable boundary conditions
\begin{equation}
  \bU={\bf0}\qquad\hbox{at}\quad x=0,\qquad\qquad
  \bU=-2Ad\,\be_y\qquad\hbox{at}\quad x=d
\end{equation}
apply.  The basic flow is the plane Couette flow,
\begin{equation}
  \bU=-2Ax\,\be_y.
\end{equation}
A modified pressure quadratic in $x$ is required to balance the
Coriolis force.

When this model is taken as a local representation of a differentially
rotating flow with angular velocity $\Omega(r)$, the shear parameter
$A$ is to be interpreted as Oort's first constant.  When the Rayleigh
discriminant $4\Omega(\Omega-A)$ is positive, the flow of an inviscid,
unmagnetized flow is linearly stable to axisymmetric perturbations.
For a Keplerian flow in which $\Omega\propto r^{-3/2}$, we have
$A/\Omega=3/4$.  (In the rheological literature, the shear rate $|2A|$
would usually be called $\dot\gamma$.)

In the case of the viscoelastic fluid, the non-zero stress components
associated with the basic flow are
\begin{equation}
  T_{xy}=T_{yx}=-2A\mu_{\rm p},\qquad\qquad
  T_{yy}=8A^2\tau\mu_{\rm p},
\end{equation}
which provide the steady solution of equation (\ref{stress}).  The
polymeric stress defined in equation (\ref{tptm}) is
\begin{equation}
  \stress_{\rm p}={{\mu_{\rm p}}\over{\tau}}
  \left[\matrix{1&-{\it De}&0\cr-{\it De}&2\,{\it De}^2+1&0\cr
  0&0&1\cr}\right],
\end{equation}
where ${\it De}=2A\tau$ is the Deborah number, and this can be
represented in the form (\ref{bb}) using the three solenoidal fields
\begin{equation}
  \bB_{1,2}=\left({{\mu_0\mu_{\rm p}}\over{2\tau}}\right)^{1/2}
  \left[\matrix{-1\cr{\it De}\pm({\it De}^2+1)^{1/2}\cr0}\right],\qquad\qquad
  \bB_3=\left({{\mu_0\mu_{\rm p}}\over{\tau}}\right)^{1/2}
  \left[\matrix{0\cr0\cr1}\right].
  \label{b123}
\end{equation}
Note that, for large ${\it De}$, the field $\bB_1$ is much greater
than the other two and corresponds to a uniform magnetic field almost
exactly in the $y$-direction.

In the case of the MHD fluid, we suppose that a uniform magnetic field
$\bB=B_y\,\be_y$ is imposed.  We also suppose the boundaries to be
perfectly conducting, so that the additional boundary conditions
\begin{equation}
  B_x={{\partial B_y}\over{\partial x}}={{\partial B_z}\over{\partial x}}=0
\end{equation}
apply at $x=0$ and $x=d$.

We note that the magnetic stress tensor in the MHD fluid resembles the
polymeric stress tensor in the viscoelastic fluid if ${\it De}$ is
large and we identify
\begin{equation}
  {{B_y^2}\over{\mu_0}}\leftrightarrow8A^2\tau\mu_{\rm p}.
  \label{identification}
\end{equation}

The energy considerations of \S~\ref{energetics} are not affected by
the rotation of the frame of reference, because the Coriolis force
does no work on the fluid.  The eigenvalues of the rate-of-strain
tensor are $(A,-A,0)$.  As the magnetic field is uniform, the maximal
growth rate of any local instability, at least in the MHD case, is
$|A|$.  Incidentally, this proves the conjecture of Balbus \& Hawley
(1992) that the magnetorotational instability, with a suitably chosen
wavevector and in the absence of dissipation, achieves the largest
possible growth rate of any local shear instability.  (In an inviscid,
unmagnetized fluid, the largest possible growth rate of Rayleigh's
inertial instability is $\sqrt{4\Omega(A-\Omega)}$.  This is always
less than or equal to $|A|$, with equality in the case $A=2\Omega$.)

\subsection{Dimensionless groups}

We introduce the kinematic viscosities $\nu=\mu/\rho$ and $\nu_{\rm
  p}=\mu_{\rm p}/\rho$.  The four dimensionless parameters of the
viscoelastic system are the Rossby number, the Reynolds number, the
Deborah number and the viscosity ratio, defined by
\begin{equation}
  {\it Ro}={{A}\over{\Omega}},\qquad
  {\it Re}={{2Ad^2}\over{\nu}},\qquad
  {\it De}=2A\tau,\qquad
  S={{\nu}\over{\nu_{\rm p}}},
\end{equation}
respectively.  For reference, the Taylor number is ${\it Ta}={\it
  Re}^2{\it Ro}^{-1}(1-{\it Ro}^{-1})$.

The four dimensionless parameters of the MHD system are the Rossby
number, the Reynolds number, the magnetic Reynolds number and the
Chandrasekhar number, defined by
\begin{equation}
  {\it Ro}={{A}\over{\Omega}},\qquad
  {\it Re}={{2Ad^2}\over{\nu}},\qquad
  {\it Rm}={{2Ad^2}\over{\eta}},\qquad
  Q={{B_y^2d^2}\over{\mu_0\rho\nu\eta}},
\end{equation}
respectively.  The identification (\ref{identification}) corresponds to
\begin{equation}
  Q\leftrightarrow{{2\,{\it Rm}\,{\it De}}\over{S}}.
\end{equation}
We can therefore quote the magnetic field strength in terms of an
effective Deborah number for the MHD system, to make a direct
comparison easier.

We are interested in comparing the behaviour of the viscoelastic and
MHD systems in the limit of large ${\it De}$ and large ${\it Rm}$.
This limit could be approached in many different ways, but we choose
to do this as one might in an ideal experiment, by keeping $\rho$,
$\nu$, $\nu_{\rm p}$, $\tau$, $\eta$ and $d$ fixed while increasing
$\Omega$, $A$ and $B_y$ together.  This means that ${\it Ro}$ and $S$
are fixed while ${\it Re}\propto{\it Rm}\propto{\it De}$ and
$Q\propto{\it De}^2$.  The elasticity, ${\it De}/{\it Re}$, is fixed
in this process.

\subsection{Linear perturbations}

We now consider small deviations from the above state, such that the
Eulerian perturbation of velocity, say, is
\begin{equation}
  \Real\left[\bu(x)\exp(st+{\rm i}k_yy+{\rm i}k_zz)\right],
\end{equation}
where $s$ is the growth rate (in general complex) and $k_y$ and $k_z$
are real wavenumber components.  Differentiation of the perturbations
with respect to $x$ will be denoted by a prime, and we define
$k^2=k_y^2+k_z^2$.

\subsection{Viscoelastic fluid}

\label{viscoelastic}

The perturbations of the viscoelastic fluid satisfy the equations
\begin{equation}
  u_x'+{\rm i}k_yu_y+{\rm i}k_zu_z=0,
  \label{ve1}
\end{equation}
\begin{equation}
  \rho\left[(s-2{\rm i}Axk_y)u_x-2\Omega u_y\right]=
  -\psi'+t_{xx}'+{\rm i}k_yt_{xy}+{\rm i}k_zt_{xz}+
  \mu(u_x''-k^2u_x),
\end{equation}
\begin{equation}
  \rho\left[(s-2{\rm i}Axk_y)u_y+2(\Omega-A)u_x\right]=
  -{\rm i}k_y\psi+t_{xy}'+{\rm i}k_yt_{yy}+{\rm i}k_zt_{yz}+
  \mu(u_y''-k^2u_y),
\end{equation}
\begin{equation}
  \rho(s-2{\rm i}Axk_y)u_z=
  -{\rm i}k_z\psi+t_{xz}'+{\rm i}k_yt_{yz}+{\rm i}k_zt_{zz}+
  \mu(u_z''-k^2u_z),
\end{equation}
\begin{equation}
  t_{xx}+\tau\left[(s-2{\rm i}Axk_y)t_{xx}-2{\rm i}T_{xy}k_yu_x\right]=
  2\mu_{\rm p}u_x',
\end{equation}
\begin{equation}
  t_{xy}+\tau\left[(s-2{\rm i}Axk_y)t_{xy}+2At_{xx}-
  T_{xy}(u_x'+{\rm i}k_yu_y)-{\rm i}T_{yy}k_yu_x\right]=
  \mu_{\rm p}(u_y'+{\rm i}k_yu_x),
\end{equation}
\begin{equation}
  t_{xz}+\tau\left[(s-2{\rm i}Axk_y)t_{xz}-{\rm i}T_{xy}k_yu_z\right]=
  \mu_{\rm p}(u_z'+{\rm i}k_zu_x),
\end{equation}
\begin{equation}
  t_{yy}+\tau\left[(s-2{\rm i}Axk_y)t_{yy}+4At_{xy}-2T_{xy}u_y'-
  2{\rm i}T_{yy}k_yu_y\right]=
  2{\rm i}\mu_{\rm p}k_yu_y,
\end{equation}
\begin{equation}
  t_{yz}+\tau\left[(s-2{\rm i}Axk_y)t_{yz}+2At_{xz}-T_{xy}u_z'-{
  \rm i}T_{yy}k_yu_z\right]=
  {\rm i}\mu_{\rm p}(k_yu_z+k_zu_y),
\end{equation}
\begin{equation}
  t_{zz}+\tau\left[(s-2{\rm i}Axk_y)t_{zz}\right]=
  2{\rm i}\mu_{\rm p}k_zu_z.
  \label{ve10}
\end{equation}
These constitute a sixth-order system of linear ODEs to be solved for
the eigenvalue $s$.  The dependent variables may be taken as
$(\psi,u_x,u_y,u_y',u_z,u_z')$.  The boundary conditions
$u_x=u_y=u_z=0$ apply at $x=0,d$.

These equations must be solved numerically in general.  However, it is
instructive to analyse further the case of unsheared or `axisymmetric'
modes ($k_y=0$), which correspond to axisymmetric modes in cylindrical
geometry.  In this case, the equations have constant coefficients and
can be combined into a single equation for $u_x$,
\begin{equation}
  \left[qs+(q\nu+\nu_{\rm p}){\rm D}\right]^2{\rm D}u_x+
  4\Omega(\Omega-A)q^2k_z^2u_x-4\Omega A\tau\nu_{\rm p}k_z^2{\rm D}u_x=0,
\end{equation}
where $q=1+\tau s$ and ${\rm D}$ is the operator
\begin{equation}
  {\rm D}=-{{\rm d^2}\over{{\rm d}x^2}}+k_z^2.
\end{equation}
This equation may be investigated analytically in an approximate way
by considering solutions of a simple trigonometric form
$u_x\propto\sin(k_xx)$, although these cannot satisfy all six physical
boundary conditions.  (In \S~\ref{numerical} below we compute the
global solutions of this equation numerically.)  The local dispersion
relation corresponding to these solutions,
\begin{equation}
  \left[qs+(q\nu+\nu_{\rm p})(k_x^2+k_z^2)\right]^2(k_x^2+k_z^2)+
  4\Omega(\Omega-A)q^2k_z^2-4\Omega A\tau\nu_{\rm p}k_z^2(k_x^2+k_z^2)=0,
  \label{axisymmetric_dispersion_relation}
\end{equation}
is a quartic equation for $s$ with real coefficients.  It can be shown
that the principle of the exchange of stabilities holds: instability
first sets in at a stationary bifurcation ($s=0$), which occurs when
the constant term passes through zero, i.e. when
\begin{equation}
  (\nu+\nu_{\rm p})^2(k_x^2+k_z^2)^3+
  4\Omega(\Omega-A)k_z^2-4\Omega A\tau\nu_{\rm p}k_z^2(k_x^2+k_z^2)=0.
\end{equation}
Suppose that Rayleigh's criterion for stability,
$4\Omega(\Omega-A)>0$, is satisfied, and that $4\Omega A>0$.  When
$\tau$ is increased from zero to a sufficiently large value, a
bifurcation occurs and axisymmetric instability ensues.  To understand
this we note that, when $\nu=0$, and in the limit $\tau\gg|s|^{-1}$
with $k_z^2\gg k_x^2$, the dispersion relation
(\ref{axisymmetric_dispersion_relation}) becomes identical to the
ideal magnetorotational dispersion relation
(\ref{magnetorotational_dispersion_relation}) for a vertical magnetic
field and vertical wavevector, provided that we identify
$B_z^2\leftrightarrow\mu_0\mu_{\rm p}/\tau$.  This is precisely what
is suggested by the field $\bB_3$ of equation (\ref{b123}).  Although
the principal analogy is with a uniform magnetic field in the
$y$-direction, such a field provides no restoring force to
axisymmetric perturbations and we see instead the effect of the much
weaker field $\bB_3$.  Therefore the axisymmetric viscoelastic
instability, which we verify numerically in \S~\ref{numerical} below,
can be understood as being analogous to a magnetorotational
instability deriving from the weak vertical field.

\subsection{MHD fluid}

\label{mhd}

The perturbations of the MHD fluid satisfy the equations
\begin{equation}
  u_x'+{\rm i}k_yu_y+{\rm i}k_zu_z=0,
  \label{mhd1}
\end{equation}
\begin{equation}
  \rho\left[(s-2{\rm i}Axk_y)u_x-2\Omega u_y\right]=
  -\psi_{\rm m}'+{\rm i}\mu_0^{-1}k_yB_yb_x+\mu(u_x''-k^2u_x),
  \label{mhd2}
\end{equation}
\begin{equation}
  \rho\left[(s-2{\rm i}Axk_y)u_y+2(\Omega-A)u_x\right]=
  -{\rm i}k_y\psi_{\rm m}+{\rm i}\mu_0^{-1}k_yB_yb_y+\mu(u_y''-k^2u_y),
  \label{mhd3}
\end{equation}
\begin{equation}
  \rho(s-2{\rm i}Axk_y)u_z=
  -{\rm i}k_z\psi_{\rm m}+{\rm i}\mu_0^{-1}k_yB_yb_z+\mu(u_z''-k^2u_z),
  \label{mhd4}
\end{equation}
\begin{equation}
  b_x'+{\rm i}k_yb_y+{\rm i}k_zb_z=0,
  \label{divb}
  \label{mhd5}
\end{equation}
\begin{equation}
  (s-2{\rm i}Axk_y)b_x=
  {\rm i}k_yB_yu_x+\eta(b_x''-k^2b_x),
  \label{bx}
  \label{mhd6}
\end{equation}
\begin{equation}
  (s-2{\rm i}Axk_y)b_y+2Ab_x=
  {\rm i}k_yB_yu_y+\eta(b_y''-k^2b_y),
  \label{mhd7}
\end{equation}
\begin{equation}
  (s-2{\rm i}Axk_y)b_z=
  {\rm i}k_yB_yu_z+\eta(b_z''-k^2b_z).
  \label{mhd8}
\end{equation}
These constitute a tenth-order system of linear ODEs to be solved for
the eigenvalue $s$.  The dependent variables may be taken as
$(\psi_{\rm m},u_x,u_y,u_y',u_z,u_z',b_y,b_y',b_z,b_z')$.  The
boundary conditions $u_x=u_y=u_z=b_y'=b_z'=0$ apply at $x=0,d$.  To
eliminate $b_x$ from the problem, differentiate equation (\ref{divb})
to find $b_x''$, then substitute into equation (\ref{bx}) to find
\begin{equation}
  (s-2{\rm i}Axk_y+\eta k^2)b_x={\rm i}k_yB_yu_x-
  \eta({\rm i}k_yb_y'+{\rm i}k_zb_z').
\end{equation}
Therefore $b_x$ is determined algebraically in terms of the dependent
variables, and can be substituted where needed.  It automatically
satisfies the boundary condition $b_x=0$ at $x=0,d$.  A difficulty
would arise if the quantity $s-2{\rm i}Axk_y+\eta k^2$ were to vanish
at any point.  As this is a complex function, it is `unlikely' that
both real and imaginary parts would vanish simultaneously.  In any
case, it could vanish only for a decaying mode, and such modes are of
no interest here.

Although the linearized equations (\ref{mhd1})--(\ref{mhd8}) for the
MHD fluid appear quite different from those of the viscoelastic fluid,
equations (\ref{ve1})--(\ref{ve10}), they can be seen to correspond in
the limits $\tau\to\infty$, $\eta\to0$ if we identify
$T_{yy}\leftrightarrow B_y^2/\mu_0$, $t_{xy}\leftrightarrow
B_yb_x/\mu_0$, $t_{yy}\leftrightarrow 2B_yb_y/\mu_0$ and
$t_{yz}\leftrightarrow B_yb_z/\mu_0$, while
$T_{xy},t_{xx},t_{xz},t_{zz}\leftrightarrow0$.

In the special case of axisymmetric modes ($k_y=0$) the velocity and
magnetic perturbations are decoupled.  The magnetic perturbation
always decays if $\eta>0$.  The remaining equations can be combined
into a single equation for $u_x$,
\begin{equation}
  (s+\nu{\rm D})^2{\rm D}u_x+4\Omega(\Omega-A)k^2u_x=0.
\end{equation}
Stability is assured if Rayleigh's criterion, $4\Omega(\Omega-A)>0$,
is satisfied.

\section{Numerical investigation}

\label{numerical}

\begin{figure}
  \centerline{\epsfbox{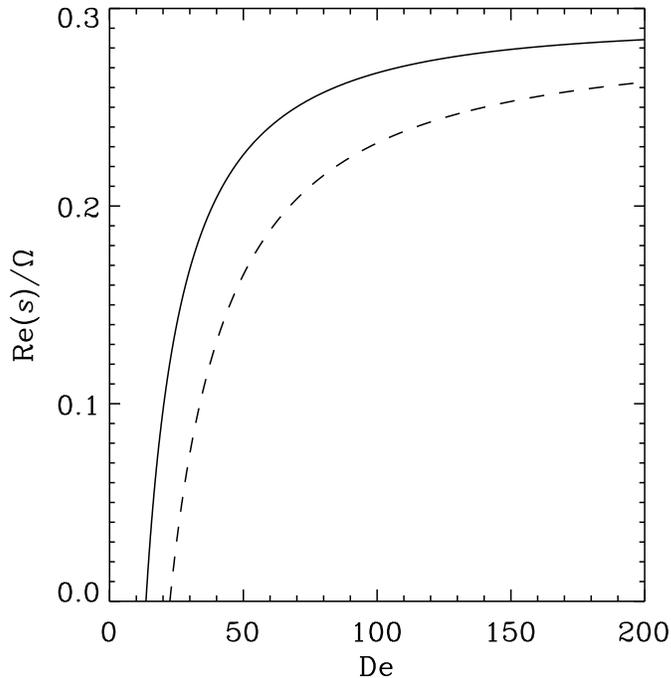}}
  \caption{Variation of the growth rate of the first unstable
    non-axisymmetric mode (at $k_y=1$) with the Deborah number, for
    the viscoelastic fluid (solid line) and the MHD fluid (dashed
    line).  Note that the Reynolds number is ${\it Re}=(24/5){\it De}$.}
\end{figure}

\begin{figure}
  \centerline{\epsfysize14cm\epsfbox{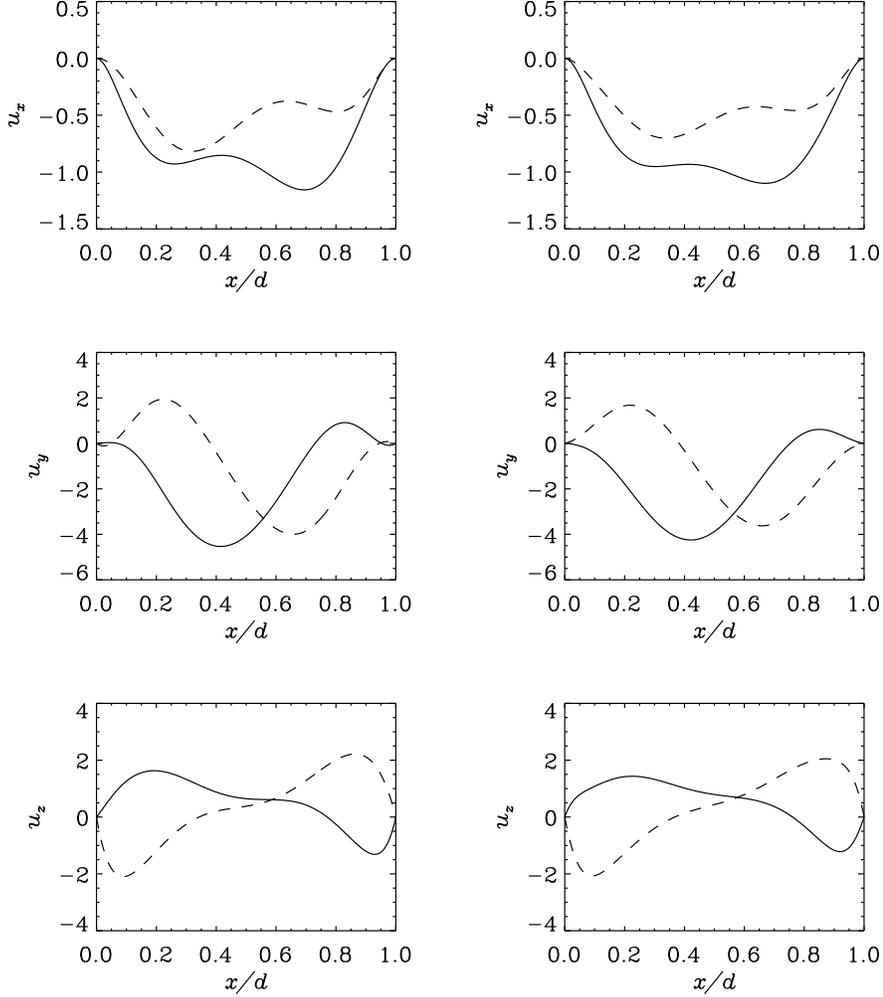}}
  \caption{Eigenfunctions of a non-axisymmetric unstable mode at
    ${\it De}=150$, for the viscoelastic fluid (left) and the MHD
    fluid (right).  Real and imaginary parts are shown with solid and
    dashed lines, respectively.}
\end{figure}

\begin{figure}
  \centerline{\epsfbox{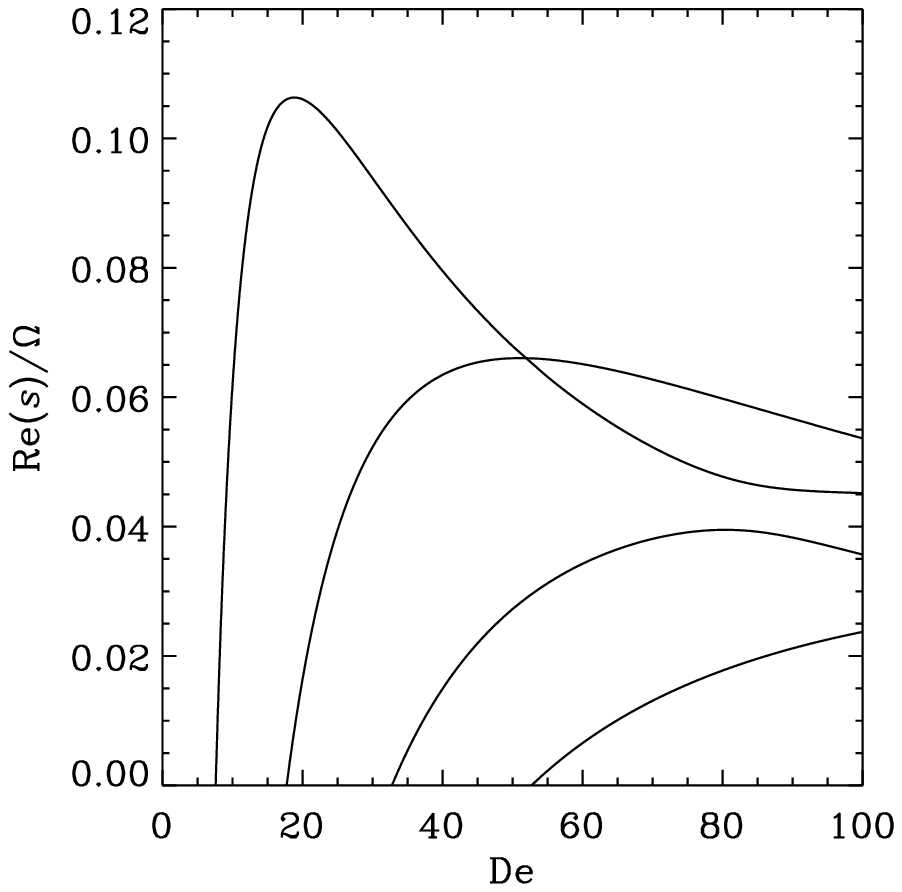}}
  \caption{Variation of the growth rates of axisymmetric unstable
    modes ($k_y=0$) with the Deborah number, for the viscoelastic
    fluid.  The eigenfunctions of modes appearing successively as
    ${\it De}$ is increased have increasing numbers of nodes.}
\end{figure}
  
We solve the eigenvalue problems defined in
\S\S~\ref{viscoelastic} and~\ref{mhd} for non-axisymmetric modes
numerically by the shooting method.  The arbitrary normalization
$\psi(0)=1$ is adopted, and the equations are integrated from $x=0$ to
$x=1$.  For the viscoelastic system, the boundary conditions at $x=1$
impose three conditions on the three unknown quantities $s$, $u_y'(0)$
and $u_z'(0)$.  Newton--Raphson iteration is applied to converge on a
solution.  For the MHD system, shooting in ${\bf C}^5$ is required.

In the absence of viscosity and resistivity, the MHD problem becomes
identical to the `Cartesian model' studied by Ogilvie \& Pringle
(1996) in their investigation of the magnetorotational instability in
the presence of an azimuthal (or toroidal) magnetic field.  We recall
some results of that analysis: (i) the instability requires a non-zero
azimuthal wavenumber $k_y$ so that the magnetic field lines are bent
by the perturbation; (ii) as $k_z$ is increased, unstable modes emerge
from the continuous spectrum of Alfv\'en waves and the eigenvalues
approach limit points; (iii) the largest growth rates are attained in
the limit $k_z\to\infty$, when the normal modes are localized near a
boundary (although, as we show below, solutions also exist that grow
rapidly but transiently in the interior of the fluid); (iv) the
maximal growth rate, $A$, is attained for an Alfv\'en frequency
$\omega_{\rm A}=(15/16)^{1/2}\Omega$ in the Keplerian case
$A/\Omega=3/4$.

For numerical purposes it is convenient to adopt $d$ and $\Omega^{-1}$
as units of length and time.  We adopt ${\it Ro}=3/4$, which is stable
according to Rayleigh's criterion and is suggested by astrophysical
applications, and take $\nu=\nu_{\rm p}=\eta$ for simplicity.

In the presence of dissipation, all modes decay in the limit
$k_z\to\infty$.  Therefore we restrict attention to a moderate value,
$k_zd=\pi$, at which the growth rates are appreciable (but not
optimal, and always less than $A$).  We select the optimal Alfv\'en
frequency, as described in \S~\ref{mri}, by choosing $k_yd=1$ and
${\it De}/{\it Re}=5/24$.  The most unstable mode is one with the
fewest nodes in its eigenfunction.  The variation of its growth rate
with ${\it De}$ is shown in figure~1.  As ${\it De}$ increases, the
eigenvalues of the mode in the viscoelastic and MHD systems converge.
The eigenfunctions are also in close agreement at ${\it De}=150$, as
shown in figure~2.

We have also solved numerically for axisymmetric unstable modes in the
viscoelastic problem, as anticipated in \S~\ref{viscoelastic}.  The
growth rates of the most unstable modes are shown in figure~3.  They
are smaller than for the non-axisymmetric modes, consistent with the
idea that the axisymmetric instability is analogous to a
magnetorotational instability deriving from the weak vertical field
$\bB_3$.

\section{Asymptotic analysis}

We now present an asymptotic analysis demonstrating the existence of
localized, non-axisymmetric growing solutions of the perturbation
equations for both systems, consistent at leading order with the
magnetorotational dispersion relation
(\ref{magnetorotational_dispersion_relation}).  We are interested
again in the limit of large Deborah number and large magnetic Reynolds
number.

\subsection{MHD fluid}

We consider a solution of the perturbation equations for the MHD
fluid, localized in a layer near an arbitrary point $x=x_0$.  Let
\begin{equation}
  x=x_0+\epsilon X,
\end{equation}
where $\epsilon\ll1$ is an ordering parameter and $X=O(1)$ within the
layer of interest.  We introduce the scalings
\begin{equation}
  k_z=\epsilon^{-3/2}\tilde k_z,\qquad
  \nu=\epsilon^4\tilde\nu,\qquad
  \eta=\epsilon^4\tilde\eta,
\end{equation}
implying that the vertical wavelength is even shorter than the width
of the layer, and that dissipation has only a weak effect on the
solution.  The solution will have an exponential time-dependence at
leading order, but we relax the assumption of a normal mode and allow
the solution to evolve freely on a long time scale captured by the
slow time coordinate $T=\epsilon t$.  This is achieved through the
replacement
\begin{equation}
  s\mapsto s_0+\epsilon{{\partial}\over{\partial T}}+O(\epsilon^2)
\end{equation}
in the perturbation equations, and we also replace
\begin{equation}
  {{\rm d}\over{{\rm d}x}}\mapsto\epsilon^{-1}{{\partial}\over{\partial X}}
\end{equation}
within the layer.  A consistent expansion scheme for the perturbations
is of the form
\begin{eqnarray}
  u_x&=&u_{x0}(X,T)+\epsilon u_{x1}(X,T)+O(\epsilon^2),\nonumber\\
  u_y&=&u_{y0}(X,T)+\epsilon u_{y1}(X,T)+O(\epsilon^2),\nonumber\\
  u_z&=&\epsilon^{1/2}\left[u_{z0}(X,T)+O(\epsilon)\right],\nonumber\\
  \psi_{\rm m}&=&\epsilon^2\left[\psi_0(X,T)+O(\epsilon)\right],\nonumber\\
  b_x&=&b_{x0}(X,T)+\epsilon b_{x1}(X,T)+O(\epsilon^2),\nonumber\\
  b_y&=&b_{y0}(X,T)+\epsilon b_{y1}(X,T)+O(\epsilon^2),\nonumber\\
  b_z&=&\epsilon^{1/2}\left[b_{z0}(X,T)+O(\epsilon)\right].
\end{eqnarray}

From equations (\ref{mhd2}), (\ref{mhd3}), (\ref{mhd6}) and
(\ref{mhd7}) at leading order we obtain the algebraic system
\begin{eqnarray}
  \rho(\hat su_{x0}-2\Omega u_{y0})&=&{\rm i}\mu_0^{-1}k_yB_yb_{x0},
  \nonumber\\
  \rho\left[\hat su_{y0}+2(\Omega-A)u_{x0}\right]&=&
  {\rm i}\mu_0^{-1}k_yB_yb_{y0},\nonumber\\
  \hat sb_{x0}&=&{\rm i}k_yB_yu_{x0},\nonumber\\
  \hat sb_{y0}+2Ab_{x0}&=&{\rm i}k_yB_yu_{y0},
\end{eqnarray}
where $\hat s=s_0-2{\rm i}Ak_yx_0$.  These may be combined into the
single equation
\begin{equation}
  \left\{\hat s^4+\hat s^2\left[4\Omega(\Omega-A)+2\omega_{\rm A}^2\right]+
  \omega_{\rm A}^2(\omega_{\rm A}^2-4\Omega A)\right\}u_{x0}=0,
  \label{ux0}
\end{equation}
where $\omega_{\rm A}^2=k_y^2B_y^2/\mu_0\rho$.  This has a
non-trivial solution,
\begin{equation}
  u_{x0}=F(X,T),
\end{equation}
if and only if $\hat s$ satisfies the magnetorotational dispersion
relation (\ref{magnetorotational_dispersion_relation}).  We then
deduce $u_{y0}$, $b_{x0}$ and $b_{y0}$ in terms of $F$, and also
$u_{z0}$, $\psi_0$ and $b_{z0}$ from equations (\ref{mhd1}),
(\ref{mhd4}) and either (\ref{mhd5}) or (\ref{mhd8}) at leading order.

From equations (\ref{mhd2}), (\ref{mhd3}), (\ref{mhd6}) and
(\ref{mhd7}) at order $\epsilon$ we similarly obtain
\begin{equation}
  \left\{\hat s^4+\hat s^2\left[4\Omega(\Omega-A)+2\omega_{\rm A}^2\right]+
  \omega_{\rm A}^2(\omega_{\rm A}^2-4\Omega A)\right\}u_{x1}=R,
  \label{ux1}
\end{equation}
where the right-hand side $R$ depends on $F$ and its derivatives.
Given that $\hat s$ has been chosen to satisfy the dispersion
relation, the solvability condition for this equation is $R=0$, which
results in an evolutionary equation for $F$,
\begin{equation}
  {{\partial F}\over{\partial T}}=a{{\partial^2F}\over{\partial X^2}}
  +({\rm i}bX-c)F.
 \label{f_equation}
\end{equation}
This is a modified diffusion equation containing constant coefficients
\begin{equation}
  a={{(\hat s+\omega_{\rm A}^2)^3}\over{2\tilde k_z^2\hat s
  \left[\hat s^4+2\omega_{\rm A}^2\hat s^2+
  \omega_{\rm A}^2(\omega_{\rm A}^2+4\Omega^2)\right]}},
\end{equation}
\begin{equation}
  b=2Ak_y,
\end{equation}
\begin{equation}
  c={{\tilde k_z^2\left[\tilde\nu(\hat s+\omega_{\rm A}^2)^2+
  4\tilde\eta\Omega^2\omega_{\rm A}^2\right]}\over
  {\hat s^4+2\omega_{\rm A}^2\hat s^2+
  \omega_{\rm A}^2(\omega_{\rm A}^2+4\Omega^2)}}.
\end{equation}
When the conditions for instability are met, $\hat s$ is real and
positive and therefore $a$, $b$ and $c$ are real and $a$ and $c$ are
positive.  A particular solution of equation (\ref{f_equation}),
corresponding to an initial condition $F(X,0)=\delta(X)$, and valid
for $T>0$ in the absence of boundaries, is the Green function
\begin{equation}
  F=(4\pi a T)^{-1/2}\exp\left[-\left(
  {{a^2b^2T^4+12acT^2-6{\rm i}abXT^2+3X^2}\over{12aT}}\right)\right],
\end{equation}
as can be obtained by Fourier-transform methods.  The Green function
decays as $T\to\infty$ for any fixed $X$, or as $|X|\to\infty$ for any
fixed $T$.

It follows that localized solutions exist that grow exponentially at
leading order, following the magnetorotational dispersion relation.
The envelope of the solution evolves more slowly in time but
ultimately decays superexponentially, so that the instability grows
for many e-folding times before the development of very short length
scales leads to decay.  If we insisted on having a normal-mode
solution, equation (\ref{f_equation}) would become an Airy equation in
$X$.  It can be shown (Ogilvie 1997) that localized solutions of this
type do exist, but only near the boundaries of the fluid.

The reason for considering disturbances that are localized in $x$ is
that it provides a convenient method of demonstrating the existence of
growing solutions without resorting to numerical analysis.  Provided
that the localization scale $\delta x$ is long compared to the
vertical wavelength and to the characteristic dissipative scales, the
growth rate is insensitive to $\delta x$.  Terquem \& Papaloizou
(1996) have shown that, in the limit of ideal MHD, a continuous
spectrum of infinitely localized growing disturbances exists.

\subsection{Viscoelastic fluid}

A very similar analysis can be carried out for the viscoelastic fluid.
The additional requirements are the scaling
$\tau=\epsilon^{-4}\tilde\tau$ and the expansions
\begin{eqnarray}
  t_{xx}&=&O(\epsilon^4),\nonumber\\
  t_{xy}&=&t_{xy0}(X,T)+\epsilon t_{xy1}(X,T)+O(\epsilon^2),\nonumber\\
  t_{xz}&=&O(\epsilon^{9/2}),\nonumber\\
  t_{yy}&=&t_{yy0}(X,T)+\epsilon t_{yy1}(X,T)+O(\epsilon^2),\nonumber\\
  t_{yz}&=&\epsilon^{1/2}\left[t_{yz0}(X,T)+O(\epsilon)\right],\nonumber\\
  t_{zz}&=&O(\epsilon^7).
\end{eqnarray}
Otherwise the analysis is so similar that we do not repeat it in
detail.  Equations (\ref{ux0}) and (\ref{ux1}) are obtained exactly as
before, provided that we identify $\omega_{\rm A}^2=k_y^2T_{yy}/\rho$
in the dispersion relation.  Exactly the same evolutionary equation
(\ref{f_equation}) is also obtained with the sole exception that the
term involving $\tilde\eta$ does not appear in the coefficient $c$.

Therefore this method also establishes the correspondence between the
viscoelastic and MHD fluids in the limit of large ${\it De}$ and large
${\it Rm}$, and demonstrates the existence of the magnetorotational
instability in the Oldroyd-B fluid.

\section{Discussion}

We have demonstrated a close analogy between a viscoelastic medium and
an electrically conducting fluid containing a magnetic field.  Both an
Oldroyd-B fluid, in the limit of large Deborah number, and a
magnetohydrodynamic fluid, in the limit of large magnetic Reynolds
number, feature a stress tensor that is nearly `frozen in' to the
fluid in a precise mathematical sense.  As a definite example of this
analogy, we have examined a local model of a differentially rotating
fluid, consisting of plane Couette flow in a rotating channel.  The
stress tensor in the case of a viscoelastic fluid resembles the
Maxwell stress corresponding to a magnetic field aligned with the
flow.

Our analysis demonstrates that there is a detailed correspondence
between instabilities in the two systems.  We have identified a direct
equivalent of the magnetorotational instability in the viscoelastic
fluid.  It exists when the angular velocity and relative vorticity are
antiparallel (or when the angular velocity decreases outwards) and the
maximal growth rate is equal to the shear parameter, or Oort constant,
$A$.  It is distinguished most clearly from Rayleigh's inertial
instability by the fact that it occurs even when the specific angular
momentum increases outwards.  It is also distinct from the elastic
instability described by Larson {\it et al.} (1990), which depends on
the curved geometry of Couette flow and exists in the elastic limit,
${\it De}/{\it Re}\to\infty$.

We have also found an axisymmetric viscoelastic instability that can
be understood as being analogous to the magnetorotational instability
of a vertical magnetic field.  This reflects the fact that the
polymeric stress tensor $\stress_{\rm p}$ can be decomposed into three
effective magnetic fields, one of which is a uniform field almost
aligned with the flow and another of which is a uniform vertical
field.  Although the vertical field is much weaker in the limit of
large ${\it De}$, it provides the dominant restoring force for
axisymmetric disturbances.

The instability discussed by Larson {\it et al.} (1990) does not
appear in our analysis because we have neglected the curvature of the
streamlines.  Although Larson {\it et al.} (1990) considered the limit
of a narrow gap, the characteristic growth rates they obtained are
smaller than the growth rates we have discussed, by a factor of order
$\epsilon^{1/2}$, where $\epsilon$ is the ratio of the gap width to
the radius.  Being purely elastic in nature, the instability of Larson
{\it et al.} (1990) must derive its energy from the elastic energy
stored in the flow, rather than the shear energy which is the source
for inertial and magnetorotational instabilities.

It is natural to enquire whether there is an MHD equivalent of the
instability of Larson {\it et al.} (1990), in which energy is derived
from the magnetic field.  In cylindrical Couette flow at large ${\it
  De}$ the dominant stress component has the form $T_{\phi\phi}\propto
r^{-4}$, which can be identified with an azimuthal (or toroidal)
magnetic field $B_\phi\propto r^{-2}$.  Typically, toroidal pinch
configurations are unstable to modes that rely on the curvature of the
magnetic field lines and derive energy from the magnetic
configuration.  The most dangerous are the $m=1$ `kink' modes and
$m=0$ `sausage' modes.  However, in the absence of fluid motion the
profile $B_\phi\propto r^{-2}$ is sufficiently steep to be stable to
all perturbations (Tayler 1973).  If the profile of $T_{\phi\phi}$
happened to be less steep, for example $T_{\phi\phi}\propto r^{-1}$,
it is likely that there would be a viscoelastic equivalent of the kink
instability.  We therefore conclude that the instability of Larson
{\it et al.} (1990) is not directly related to an MHD instability, but
relies on inherently viscoelastic effects not captured by our analogy.
This conclusion is supported by an examination of the physical
explanation that Larson {\it et al.} give for their instability.

Since the work of Larson {\it et al.} (1990) there have been a number
of related theoretical and experimental studies of viscoelastic
Couette flow, some of which have examined the effects of inertia and
non-axisymmetry (e.g. Avgousti \& Beris 1993; Steinberg \& Groisman
1998; Baumert \& Muller 1999).  Some of these might have been expected
to reveal the analogue of the magnetorotational instability.  However,
it appears that the cases usually investigated are those in which
either the outer or inner cylinder is stationary, or the inner
cylinder rotates at twice the angular velocity of the outer cylinder
with only a narrow gap between the two.  This means that, whenever the
analogue of the magnetorotational instability might have occurred, the
system is unstable to Rayleigh's inertial instability.  In order to
separate the two effects, it would be valuable to examine cases in
which the angular velocity decreases outwards but the specific angular
momentum increases.  Interestingly, the magnetorotational instability
has never been demonstrated in laboratory experiments.  Although there
is currently much effort towards this goal (e.g. Goodman \& Ji 2002),
the technical requirements are considerable and the system is
constrained by the very small magnetic Prandtl numbers of liquid
metals.  A viscoelastic magnetorotational experiment might prove to be
less demanding and easier to visualize.

We conclude with some further perspectives on the analogy between
viscoelastic and MHD flows.

The analogy is asymptotic in nature and therefore not perfect.
Viscoelastic and MHD flows deviate from simple stress freezing in
different ways: the viscoelastic stress relaxes, while the magnetic
field diffuses.  The classes of exactly steady solutions of the two
systems are therefore different, because after a sufficiently long
time either relaxation or diffusion will have its effect.  For
example, while there is only one solution for viscoelastic Couette
flow, magnetized Couette flow can be set up with either vertical or
azimuthal current-free magnetic fields.  In this sense, the analogy is
more applicable to dynamical, time-dependent situations than to steady
flows.

Renardy (1997) has analysed the large-${\it De}$ limit of steady,
two-dimensional flows of the upper-convected Maxwell fluid (obtained
by setting the solvent viscosity $\mu$ of the Oldroyd-B fluid to
zero).  Noting that the stress $\stress$ has one dominant eigenvalue
in this limit, he writes $\stress=\rho\bu\bu$, where $\rho$ and $\bu$
are a fictitious density and a fictitious velocity field.  When
inertial forces are negligible, $\bnabla\bcdot\stress$ must balance
the pressure gradient and $\rho$ and $\bu$ are then found to satisfy
the steady Euler equations.  Through our analogy, a connection can be
seen here with the work of Moffatt (1985), which makes use of the
analogy between steady Euler flows and magnetostatic equilibria in
which the Lorentz force balances the pressure gradient.

There is a close connection between the polymeric stress $\stress_{\rm
  p}$ at {\it moderate\/} De and the mean stress tensor
$\langle\bB\bB\rangle/\mu_0$ of a {\it disordered\/} magnetic field,
as occurs in MHD turbulence.  Under these conditions, $\stress_{\rm
  p}$ and $\langle\bB\bB\rangle/\mu_0$ may have three positive
eigenvalues of comparable magnitude.  Indeed, Ogilvie (2001) has
suggested the use of Maxwellian viscoelastic models, with Deborah
numbers of order unity, for MHD turbulence in accretion discs.

Besides Couette flow, another problem that has received much attention
is the stability of a planar jet or shear layer with respect to
two-dimensional disturbances.  Azaiez \& Homsy (1994) and Rallison \&
Hinch (1995) examined the equivalent problem for a viscoelastic fluid,
noting the potentially stabilizing influence of a polymer additive.
Taking a limit in which ${\it Re}$ and ${\it De}$ tend to infinity
while maintaining a finite ratio, they derived an elastic equivalent
of Rayleigh's stability equation.  An analogy exists between this
problem and that of the stability of a similar flow of an ideal,
electrically conducting fluid with a magnetic field parallel to the
flow, a problem studied since the 1950s.  In a recent study, Hughes \&
Tobias (2001) derived a magnetic Rayleigh equation (their equation
(3.5)) that is exactly equivalent to the elastic Rayleigh equation
given by Rallison \& Hinch (1995, p.~314).

In this paper we have drawn attention to a useful analogy between
viscoelastic and MHD flows, and have discussed the relation between
instabilities of differential rotation in the two systems.  We
anticipate that much further use can be made of this analogy.

\begin{acknowledgments}
  We are grateful to John Hinch, John Rallison and the referees of
  this paper for their constructive comments.  G.\,I.\,O. acknowledges
  the support of the Royal Society through a University Research
  Fellowship.
\end{acknowledgments}


\begin{thebibliography}{}

\bibitem[Alfv\'en (1950)]{A50}
  \textsc{Alfv\'en, H.} 1950
  {\it Cosmical Electrodynamics.}  Oxford University Press.

\bibitem[Avgousti \& Beris (1993)]{AB93}
  \textsc{Avgousti, M. \& Beris, A. N.} 1993
  Non-axisymmetric modes in viscoelastic Taylor--Couette flow.
  \textit{J. Non-Newtonian Fluid Mech.} \textbf{50}, 225--251.

\bibitem[Azaiez \& Homsy (1994)]{AH94}
  \textsc{Azaiez, J. \& Homsy, G. M.} 1994
  Linear stability of free shear flow of viscoelastic liquids.
  \textit{J. Fluid Mech.} \textbf{268}, 37--69.

\bibitem[Balbus \& Hawley (1991)]{BH91}
  \textsc{Balbus, S. A. \& Hawley, J. F.} 1991
  A powerful local shear instability in weakly magnetized disks.
  I. Linear analysis.
  \textit{Astrophys. J.} \textbf{376}, 214--233.

\bibitem[Balbus \& Hawley (1992)]{BH92}
  \textsc{Balbus, S. A. \& Hawley, J. F.} 1992
  Is the Oort $A$-value a universal growth rate limit for accretion disk
  shear instabilities?
  \textit{Astrophys. J.} \textbf{392}, 662--666.

\bibitem[Balbus \& Hawley (1998)]{BH98}
  \textsc{Balbus, S. A. \& Hawley, J. F.} 1998
  Instability, turbulence, and enhanced transport in accretion disks.
  \textit{Rev. Mod. Phys.} \textbf{70}, 1--53.

\bibitem[Baumert \& Muller (1999)]{BM99}
  \textsc{Baumert, B. M. \& Muller, S. J.} 1999
  Axisymmetric and non-axisymmetric elastic and inertio-elastic
  instabilities in Taylor--Couette flow.
  \textit{J. Non-Newtonian Fluid Mech.} \textbf{83}, 33--69.

\bibitem[Bird, Armstrong \& Hassager (1987)]{BAH87}
  \textsc{Bird, R. B., Armstrong, R. C. \& Hassager, O.} 1987{\it a\/}
  {\it Dynamics of Polymeric Liquids}, 2nd Edn, vol. 1.  Wiley.

\bibitem[Bird, Curtiss \& Armstrong \& Hassager (1987)]{BCA87}
  \textsc{Bird, R. B., Curtiss, C. F. \& Armstrong, R. C.} 1987{\it b\/}
  {\it Dynamics of Polymeric Liquids}, 2nd Edn, vol. 2.  Wiley.

\bibitem[Chandrasekhar (1960)]{C60}
  \textsc{Chandrasekhar, S.} 1960
  The stability of non-dissipative Couette flow in hydromagnetics.
  \textit{Proc. Natl Acad. Sci.} \textbf{46}, 253--257.

\bibitem[Chandrasekhar (1961)]{C61}
  \textsc{Chandrasekhar, S.} 1961
  {\it Hydrodynamic and Hydromagnetic Stability.}  Oxford University Press.

\bibitem[Clerk Maxwell (1867)]{CM67}
  \textsc{Clerk Maxwell, J.} 1867
  On the dynamical theory of gases.
  \textit{Phil. Trans.} \textbf{157}, 49--88.
  Reprinted in \textit{Scientific Papers of James Clerk Maxwell},
  vol. 2 (ed.\ W. D. Niven), p.\ 26.  Cambridge University Press.

\bibitem[Giesekus (1966)]{G66}
  \textsc{Giesekus, H.} 1966
  Zur Stabilt\"at von Str\"omungen viskoelastischer Fl\"ussigkeiten.
  1. Ebene und kreisf\"ormige Couette-Str\"omung.
  \textit{Rheologica Acta} \textbf{5}, 239--252.

\bibitem[Goodman \& Ji (2002)]{GJ02}
  \textsc{Goodman, J. \& Ji, H.} 2002
  Magnetorotational instability of dissipative Couette flow.
  \textit{J. Fluid Mech.} \textbf{462}, 365--382.

\bibitem[Hughes \& Tobias (2001)]{HT01}
  \textsc{Hughes, D. W. \& Tobias, S. M.} 2001
  On the instability of magnetohydrodynamic shear flows.
  \textit{Proc. R. Soc. Lond.} A \textbf{457}, 1365--1384.

\bibitem[Joseph (1990)]{J90}
  \textsc{Joseph, D. D.} 1990
  {\it Fluid Dynamics of Viscoelastic Liquids.}  Springer.

\bibitem[Larson, Shaqfeh \& Muller (1990)]{LSM90}
  \textsc{Larson, R. G., Shaqfeh, E. S. G. \& Muller, S. J.} 1990
  A purely elastic instability in Taylor--Couette flow.
  \textit{J. Fluid Mech.} \textbf{218}, 573--600.

\bibitem[Moffatt (1985)]{M85}
  \textsc{Moffatt, H. K.} 1985
  Magnetostatic equilibria and analogous Euler flows of arbitrarily
  complex topology.  Part 1. Fundamentals.
  \textit{J. Fluid Mech.} \textbf{159}, 359--378.

\bibitem[Ogilvie (2001)]{O97}
  \textsc{Ogilvie, G. I.} 1997
  Magnetic fields in accretion discs.
  PhD dissertation, University of Cambridge.

\bibitem[Ogilvie (2001)]{O01}
  \textsc{Ogilvie, G. I.} 2001
  Non-linear fluid dynamics of eccentric discs.
  \textit{Mon. Not. R. Astron. Soc.} \textbf{325}, 231--248.

\bibitem[Ogilvie \& Pringle (1996)]{OP96}
  \textsc{Ogilvie, G. I. \& Pringle, J. E.} 1996
  The non-axisymmetric instability of a cylindrical shear flow
  containing an azimuthal magnetic field.
  \textit{Mon. Not. R. Astron. Soc.} \textbf{279}, 152--164.

\bibitem[Oldroyd (1950)]{O50}
  \textsc{Oldroyd, J. G.} 1950
  On the formulation of rheological equations of state.
  \textit{Proc. R. Soc. Lond.} A \textbf{200}, 523--541.

\bibitem[Priest (1982)]{P82}
  \textsc{Priest, E. R.} 1982
  {\it Solar Magnetohydrodynamics.}  Reidel.

\bibitem[Pringle (1981)]{P81}
  \textsc{Pringle, J. E.} 1981
  Accretion discs in astrophysics.
  \textit{Annu. Rev. Astron. Astrophys.} \textbf{19}, 137--162.

\bibitem[Rallison \& Hinch (1995)]{RH95}
  \textsc{Rallison, J. M. \& Hinch, E. J.} 1995
  Instability of a high-speed submerged elastic jet.
  \textit{J. Fluid Mech.} \textbf{288}, 311--324.

\bibitem[Rayleigh (1916)]{R16}
  \textsc{Rayleigh, Lord} 1916
  On the dynamics of revolving fluids.
  \textit{Proc. R. Soc. Lond.} A \textbf{93}, 148--154.

\bibitem[Renardy (1997)]{R97}
  \textsc{Renardy, M.} 1997
  The high Weissenberg number limit of the UCM model and the Euler equations.
  \textit{J. Non-Newtonian Fluid Mech.} \textbf{69}, 293--301.

\bibitem[Roberts (1967)]{R67}
  \textsc{Roberts, P. H.} 1967
  {\it An Introduction to Magnetohydrodynamics.}  Longmans.

\bibitem[Steinberg \& Groisman (1998)]{SG98}
  \textsc{Steinberg, V. \& Groisman, A.} 1998
  Elastic versus inertial instability in Couette--Taylor flow of a
  polymer solution: review.
  \textit{Phil. Mag.} B \textbf{78}, 253--263.

\bibitem[Tayler (1973)]{T73}
  \textsc{Tayler, R. J.} 1973
  The adiabatic stability of stars containing magnetic fields ---
  I. Toroidal fields.
  \textit{Mon. Not. R. Astron. Soc.} \textbf{161}, 365--380.

\bibitem[Taylor (1923)]{T23}
  \textsc{Taylor, G. I.} 1923
  Stability of a viscous liquid contained between two rotating cylinders.
  \textit{Phil. Trans. R. Soc. Lond.} A \textbf{223}, 289--343.

\bibitem[Terquem \& Papaloizou (1996)]{TP96}
  \textsc{Terquem, C. \& Papaloizou, J. C. B.} 1996
  On the stability of an accretion disc containing a toroidal magnetic field.
  \textit{Mon. Not. R. Astron. Soc.} \textbf{279}, 767--784.

\bibitem[Thomas \& Walters (1964)]{TW64}
  \textsc{Thomas, R. H. \& Walters, K.} 1964
  The stability of elastico-viscous flow between rotating cylinders.  Part 1.
  \textit{J. Fluid Mech.} \textbf{18}, 33--43.

\bibitem[Tur \& Yanovsky (1993)]{TV93}
  \textsc{Tur, A. V. \& Yanovsky, V. V.} 1993
  Invariants in dissipationless hydrodynamic media.
  \textit{J. Fluid Mech.} \textbf{248}, 67--106.

\bibitem[Velikhov (1959)]{V59}
  \textsc{Velikhov, E. P.} 1959
  Stability of an ideally conducting liquid flowing between cylinders
  rotating in a magnetic field.
  \textit{Sov. Phys. JETP} \textbf{9}, 995--998.

\end{thebibliography}
\end{document}